\documentclass[aip,apl,unsortedaddress,superscriptaddress,reprint]{revtex4-1}
\usepackage{hyperref}
\usepackage{graphicx}
\usepackage{amsmath}
\usepackage{amssymb}
\usepackage{times,mathptmx}
\usepackage{dcolumn}
\usepackage{ifthen}
\usepackage{color}
\usepackage{proof}

\begin{document}

\title[]{Effect of interfacial oxidation layer in spin pumping experiments on Ni$_{80}$Fe$_{20}$/SrIrO$_3$ heterostructures}
\affiliation{Walther-Mei{\ss}ner-Institut, Bayerische Akademie der Wissenschaften, 85748 Garching, Germany}
\affiliation{Department of Physics, Nano Functional Materials Technology Center, Material Science Research Center, Indian Institute of Technology Madras (IITM), India 600036}
\affiliation{Low Temperature Physics Lab, Department of Physics, Indian Institute of Technology Madras (IITM), Chennai, India 600036}
\affiliation{Physik-Department, Technische Universit\"at M\"unchen, 85748 Garching, Germany}
\affiliation{Munich Center for Quantum Science and Technology (MCQST), Schellingstr. 4, 80799 M\"unchen, Germany}

\author{T.S. Suraj}
\email{surajts@physics.iitm.ac.in}
\affiliation{Department of Physics, Nano Functional Materials Technology Center, Material Science Research Center, Indian Institute of Technology Madras (IITM), India 600036}
\affiliation{Low Temperature Physics Lab, Department of Physics, Indian Institute of Technology Madras (IITM), Chennai, India 600036}
\author{Manuel M\"uller}%
\affiliation{Walther-Mei{\ss}ner-Institut, Bayerische Akademie der Wissenschaften, 85748 Garching, Germany}
\author{Sarah Gelder}
\affiliation{Walther-Mei{\ss}ner-Institut, Bayerische Akademie der Wissenschaften, 85748 Garching, Germany}
\author{Stephan Gepr\"ags}
\affiliation{Walther-Mei{\ss}ner-Institut, Bayerische Akademie der Wissenschaften, 85748 Garching, Germany}
\author{Matthias Opel}
\affiliation{Walther-Mei{\ss}ner-Institut, Bayerische Akademie der Wissenschaften, 85748 Garching, Germany}
\author{Mathias Weiler}
\affiliation{Walther-Mei{\ss}ner-Institut, Bayerische Akademie der Wissenschaften, 85748 Garching, Germany}
\affiliation{Physik-Department, Technische Universit\"at M\"unchen, 85748 Garching, Germany}
\author{K. Sethupathi}
\affiliation{Low Temperature Physics Lab, Department of Physics, Indian Institute of Technology Madras (IITM), Chennai, India 600036}
\author{Hans Huebl}
\affiliation{Walther-Mei{\ss}ner-Institut, Bayerische Akademie der Wissenschaften, 85748 Garching, Germany}
\affiliation{Physik-Department, Technische Universit\"at M\"unchen, 85748 Garching, Germany}
\affiliation{Munich Center for Quantum Science and Technology (MCQST), Schellingstr. 4, 80799 M\"unchen, Germany}
\author{Rudolf Gross}
\affiliation{Walther-Mei{\ss}ner-Institut, Bayerische Akademie der Wissenschaften, 85748 Garching, Germany}
\affiliation{Physik-Department, Technische Universit\"at M\"unchen, 85748 Garching, Germany}
\affiliation{Munich Center for Quantum Science and Technology (MCQST), Schellingstr. 4, 80799 M\"unchen, Germany}
\author{M.S. Ramachandra Rao}
\affiliation{Department of Physics, Nano Functional Materials Technology Center, Material Science Research Center, Indian Institute of Technology Madras (IITM), India 600036}
\author{Matthias Althammer}
\email{matthias.althammer@wmi.badw.de}
\affiliation{Walther-Mei{\ss}ner-Institut, Bayerische Akademie der Wissenschaften, 85748 Garching, Germany}
\affiliation{Physik-Department, Technische Universit\"at M\"unchen, 85748 Garching, Germany}

\date{\today}

\begin{abstract}
SrIrO$_3$ with its large spin-orbit coupling and low charge conductivity has emerged as a potential candidate for efficient spin-orbit torque magnetization control in spintronic devices. We here report on the influence of an interfacial oxide layer on spin pumping experiments in Ni$_{80}$Fe$_{20}$ (NiFe)/SrIrO$_3$ bilayer heterostructures. To investigate this scenario we have carried out broadband ferromagnetic resonance (BBFMR) measurements, which indicate the presence of an interfacial antiferromagnetic oxide layer. We performed in-plane BBFMR experiments at cryogenic temperatures, which allowed us to simultaneously study dynamic spin pumping properties (Gilbert damping) and static magnetic properties (such as the effective magnetization and magnetic anisotropy). The results for NiFe/SrIrO$_3$ bilayer thin films were analyzed and compared to those from a NiFe/NbN/SrIrO$_3$ trilayer reference sample, where a spin-transparent, ultra-thin NbN layer was inserted to prevent oxidation of NiFe. At low temperatures, we observe substantial differences in the magnetization dynamics parameters of these samples, which can be explained by an antiferromagnetic interfacial layer in the NiFe/SrIrO$_3$ bilayers.
\end{abstract}
\maketitle

Charge to spin current conversion efficiency in heavy metal(HM)/ferromagnet(FM) bilayers has become one of the central themes of spintronics research, with the goal to manipulate the magnetization in the FM via spin-orbit torques (SOT) induced at the interface to the HM.~\cite{1,2,3,5,37} Heavy metals, like Pt, W and Ta, have been successfully used in SOT experiments in HM/FM bilayers owing to their large spin-orbit coupling (SOC).~\cite{4,5,6} Beyond these well-established HM materials, iridium-based oxides (iridates) with their high spin Hall conductivity, low charge conductivity and large SOC are promising candidates for SOT studies.~\cite{7,8} Among them, the 5d transition metal oxide SrIrO$_3$ (SIO) in particular remains in spotlight due to its exotic band structure with extended 5d orbitals. Compared to metals, the SOT effects in oxide materials offer a wide tunability due to the dependency of the electronic properties on the oxygen octahedral rotation or oxygen vacancies.~\cite{9}
A large spin Hall angle of 1.1 was reported for SIO from second-harmonic Hall measurements in NiFe/SIO.~\cite{12}

The study of direct SOT effects requires patterning processes and also demands sophisticated measurement protocols. On the other hand, due to Onsager reciprocity\cite{39}, spin pumping experiments allow to study the inverse SOT effects in blanket HM/FM bilayer structures. Here, magnetization dynamics is excited in the FM via an external microwave driving field and excess angular momentum is pumped as a pure spin current across the interface into the HM. Absorption of this pure spin current in the HM represents an additional contribution to the damping of the magnetization dynamics. Evaluating the change in linewidth in combination with the voltage generated by the inverse SOT enables one to address and quantitatively analyze the inverse SOT effects in the HM.~\cite{10,11} Recent spin pumping experiments demonstrated a large SOT for SIO, as compared to elemental heavy metals.~\cite{12} All the spin pumping experiments so far carried out in SIO employed the metallic ferromagnet NiFe as a spin injector layer due to its low damping, making it an ideal candidate for ferromagnetic resonance (FMR) experiments.~\cite{9} Yuelei Zhao \textit{et.al,} determined the Gilbert damping of NiFe/Al$_2$O$_3$ heterostructures as a function of temperature and observed a peak in Gilbert damping near $T\sim50\;\mathrm{K}$, which gradually becomes broader with increase in NiFe thickness and vanishes above $20\;\mathrm{nm}$ of NiFe.~\cite{13} They interpreted this as a spin reorientation of the surface magnetization of NiFe thin films arising from thermal excitations.~\cite{13} However, recently it was found that the choice of NiFe in combination with oxide substrates invoked interfacial oxidation showing antiferromagnetic ordering at low temperatures.~\cite{14} This has not been taken into account in previous experiments with NiFe/SIO bilayers, but needs to be addressed to potentially tune the SOT efficiency.

In this letter, we study the influence of an interfacial oxide layer between NiFe and SIO on spin pumping experiments. To this end, we explore the spin transport in NiFe/SIO bilayers with and without inserting a thin NbN spacer layer between SIO and NiFe. The additional spacer layer allows to prevent the diffusion of oxygen to the NiFe layer, while not strongly suppressing spin transport. We employed the BBFMR technique to study the magnetization dynamics and extracted the FMR spectroscopic parameters as a function of temperature. A comparison of these parameters for samples with and without NbN spacer layer permits us to identify the potential formation of an interfacial oxide layer.

\begin{figure}
\includegraphics[width=85mm]{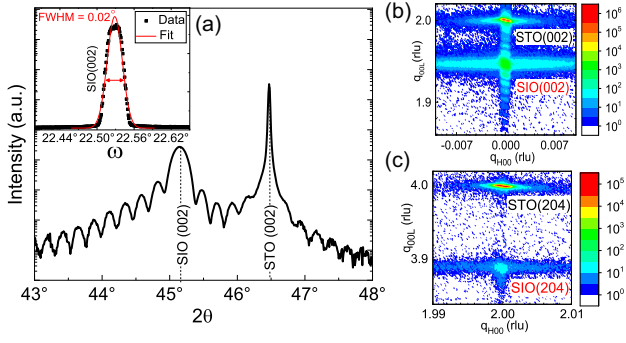}
\caption{\label{Fig1}Structural properties of a SIO ($30\;\mathrm{nm}$) thin film grown on a (001)-oriented SrTiO$_3$ (STO) substrate. (a) $2\theta$-$\omega$ scan along the [001] direction of STO. The inset shows the rocking curve of the SIO(002) reflection and the derived full width at half maximum (FWHM) value. (b), (c) Reciprocal space mappings around the symmetric STO(002) and the asymmetric STO(204) reflections, respectively. The reciprocal lattice units (rlu) are related to the respective STO(001) substrate reflection.}
\end{figure}

SIO thin films with a thickness of $5\;\mathrm{nm}$ were deposited on single crystalline, (001)-oriented SrTiO$_3$ (STO) substrates using pulsed laser deposition. \footnote{The setup is described elsewhere~\cite{15}, the growth parameters are compiled in the supplementary material.} Subsequently, NiFe (Ni$_{80}$Fe$_{20}$) was DC sputter-deposited ex-situ on top of SIO and capped with a $3\;\mathrm{nm}$ thin Al layer to prevent the top surface of NiFe from oxidation. For comparison, an additional trilayer sample was fabricated with a $3\;\mathrm{nm}$ thin NbN layer between SIO and NiFe to prevent the oxidation of the NiFe layer at this interface. NbN is a well established diffusion barrier for oxygen \cite{38}, and a superconductor with $T_\mathrm{C}$ around $18\;\mathrm{K}$\cite{29}. 
In addition, we also fabricated a NiFe thin film grown directly on top of an STO substrate and capped with $3\;\mathrm{nm}$ of Al to explore its intrinsic properties. All sputter deposition processes were performed at room temperature in an ultrahigh vacuum system (base pressure in the 10$^{-9}$ mbar range). The sputtering process was carried out at 5x10$^{-3}$ mbar in an Ar (NiFe, Al) or an Ar and N$_2$ mixture (NbN, flow ratio of Ar to N$_2$: 18.1/1.9) atmosphere. To ensure the sample quality, we performed X-ray diffraction studies and magnetometry measurements (SQUID magnetometer). Broadband ferromagnetic resonance (BBFMR) measurements employed a vector network analyzer (VNA) in combination with a 3D-vector magnet cryostat with a variable temperature insert.

High-resolution X-ray diffraction measurements reveal an epitaxial growth of SIO on STO (see Fig.~\ref{Fig1}). The high crystalline quality of the samples is confirmed by  $2\theta$ - $\omega$ scans revealing satellites around the SIO(002) reflection due to Laue oscillations, indicating a coherent growth (Fig.~\ref{Fig1}(a)). The thin films show a low mosaic spread, as demonstrated by the full width at half maximum (FWHM) value of $0.02^\circ$ extracted from the SIO(002) rocking curve  (inset in Fig.~\ref{Fig1}(a)). In addition, the reciprocal space maps (RSM) around the STO (002) (Fig.~\ref{Fig1}(b)) and STO (204) reflections (Fig.~\ref{Fig1}(c)) reveal a lattice matched growth of the SIO film on the STO as both share the same reciprocal lattice units q$_H$. Thus, the SIO exhibits a compressive strain in the film plane.

\begin{figure}
\includegraphics[width=85mm]{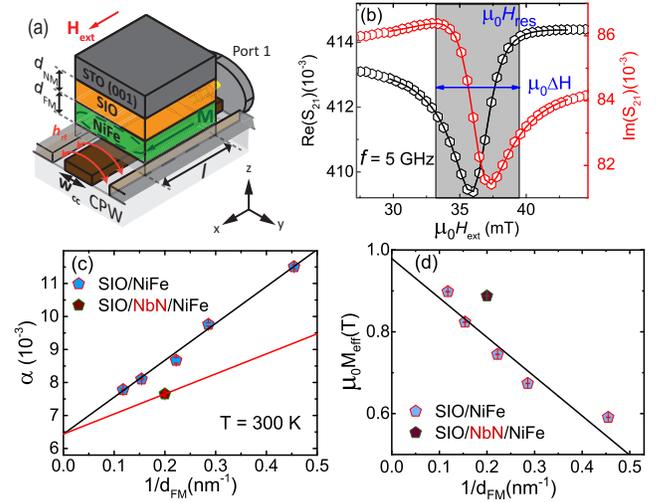}
\caption{\label{Fig2}Room temperature BBFMR measurements. (a) Schematic of the experimental set-up and illustration of the sample stack mounting on a CPW. (b) Experimental data (symbols) of the real (black) and imaginary part (red) of S$_{21}$ from the SIO/NiFe($5\;\mathrm{nm}$) sample, excited with $f=5\;\mathrm{GHz}$. The lines are fits according to Ref.~\onlinecite{22}. (c) Gilbert damping parameter, $\alpha$ as a function of $1/d_\mathrm{FM}$ in the in-plane geometry with a linear fit (black line) indicating the intrinsic damping of NiFe with spin pumping contribution for SIO/NiFe heterostructures. For comparison, the single data point of one SIO/NbN/NiFe($5\;\mathrm{nm}$) trilayer sample has been added. For the trilayer, we connected with a red line the singular data point with the same bulk damping $\alpha_0$. (d) Effective magnetization, $\mathrm{M_{eff}}$ as a function of $1/d_\mathrm{FM}$ with linear fit (line) indicating the presence of interface anisotropy in the SIO/NiFe heterostructures. For comparison, the single data point of one SIO/NbN/NiFe($5\;\mathrm{nm}$) sample is also added.}
\end{figure}

For the investigation of the magnetization dynamics, we performed BBFMR measurements using a coplanar waveguide (CPW) with a $80\;\mathrm{\mu m}$ wide center conductor. We record the complex microwave transmission parameter S$_{21}$ at fixed microwave frequencies $f$ in the range from $5\;\mathrm{GHz}$ to $32\;\mathrm{GHz}$ as a function of the in-plane magnetic field $\mu_0 H_\mathrm{ext}$ with a VNA output power of $0\;\mathrm{dBm}$. A schematic illustration of the experimental set-up is shown in Fig.~\ref{Fig2}(a). The real and imaginary part (black and red symbols) of the recorded transmission parameter S$_{21}$  are fitted (black and red lines) to the Polder susceptibility as show in Fig.~\ref{Fig2}(b).~\cite{16} From this fit we extract the values of the FMR field $\mu_{0}H_\mathrm{res}$ and the FMR linewidth $\mu_{0}\Delta H $ for each frequency. Using the Kittel formula\citep{30} for the in-plane magnetization case
\begin{equation}
    \mu_{0}H_\mathrm{res}= -\mu_{0}H_\mathrm{ani}-\frac{\mu_{0}M_\mathrm{eff}}{2}+\sqrt{{\left(\frac{\mu_{0}M_\mathrm{eff}}{2}\right)^2}+{\left(\frac{2\pi f}{\gamma}\right)^2}},
    \label{eq1}
\end{equation}
with $\gamma$ the gyromagnetic ratio. Thus, we can extract the effective saturation magnetization $M_\mathrm{eff}=M_\mathrm{s}-K_\mathrm{u}$ (with the saturation magnetization $M_\mathrm{s}$ and the out-of-plane anisotropy field $K_\mathrm{u}$) and the in-plane anisotropy field $\mu_{0}H_\mathrm{ani}$ along the CPW direction. In addition, we determine the Gilbert damping parameter $\alpha$ as well as the inhomogeneous line broadening $\mu_0 H_\mathrm{inh}$ from the microwave frequency dependence of $\mu_{0}\Delta H$ via the relation\cite{34,35}
\begin{equation}
    \mu_{0}\Delta H= \mu_{0}H_\mathrm{inh}+2\frac{2\pi f\alpha}{\gamma}
    \label{eq2}
\end{equation}
(see supplementary material for frequency dependent BBFMR data).

To quantify the role of spin pumping in our SIO($5\;\mathrm{nm}$)/NiFe($d_\mathrm{FM}$) bilayer heterostructures, we extracted the Gilbert damping parameter $\alpha$ as a function of the NiFe layer thickness $d_\mathrm{FM}$ (see Fig.~\ref{Fig2}(c)). A linear dependence of $\alpha$ on $1/d_\mathrm{FM}$ is clearly evident and is attributed to pumping a spin current from the NiFe into SIO.~\cite{17}
Moreover, the $y$-axis intercept allows to quantify the bulk Gilbert damping $\alpha_0$, while the slope allows to determine the effective spin mixing conductance $g^{{\uparrow\downarrow}}_\mathrm{eff}$ via\cite{36}
 \begin{equation}
    \alpha(d_\mathrm{FM})= \alpha_0 + \frac{\gamma \hbar g^{{\uparrow\downarrow}}_\mathrm{eff}}{4\pi M_\mathrm{s}}\left(\frac{1}{d_\mathrm{FM}}\right).
    \label{eq4}
\end{equation}
Here, $\hbar$ is the reduced Planck's constant, and $M_\mathrm{s}= 630\;\mathrm{kA/m}$ is determined from SQUID magnetometry of our NiFe layers. We obtain $\alpha_0= 6.44\times10 ^{-3}\pm 8 \times 10^{-5}$, which corresponds well to literature.~\cite{18,19,20,21} In addition, we find  $g^{{\uparrow\downarrow}}_\mathrm{eff} = 4.68 \times 10^{18} \pm 2.2 \times 10^{17}\mathrm{m^{-2}}$ for these heterostructures at room temperature.~\cite{22,40} This result agrees well with values in literature for NiFe/heavy metal heterostructures\cite{23,24,weiler_spin_2013} and highlights the feasibility of SIO for SOT devices.

To investigate the role of an interfacial NiFeO$_x$ oxide layer, we conducted BBFMR measurements on a SIO/NbN/NiFe trilayer sample. To obtain an estimate for spin pumping from this singular sample, we note that NiFe can be grown on NbN with bulk properties (see supplementary material) and consequently connected the measured data point with a line through the singular data point with the same bulk damping $\alpha_0 = 6.44 \times 10^{-3}$.  Despite the expected long spin diffusion length of NbN ($14\;\mathrm{nm}$)~\cite{28}, we observe dramatically reduced spin pumping in this sample. Assuming the formation of an interfacial oxide NiFeO$_x$ layer when SIO is in direct contact with NiFe, our results suggest that either this oxide layer allows for more efficient spin current injection as already reported for HMs in contact with NiFe~\cite{41} or serves as a source of spin memory loss~\cite{31}. However, more systematic studies are required to unambiguously separate these two contributions and better understand the role of the interfacial NiFeO$_x$ layer for spin current transport.

The nature of the interfacial oxide layer in our bilayers were further investigated by plotting $\mu_{0}M_\mathrm{eff}$ versus $1/d_\mathrm{FM}$ as shown in Fig.~\ref{Fig2}(d). This linear scaling with $1/d_\mathrm{FM}$ is an indication of interfacial anisotropy in the bilayer sample and may be induced via the oxidation of NiFe.~\cite{42} Interestingly, for the SIO/NbN/NiFe trilayer we find a larger ($\sim$ $\mathrm{15}\%)$ $\mu_{0}M_\mathrm{eff}$ as compared to bilayer samples with similar $d_\mathrm{FM}$. This clearly indicates a change in interface anisotropy.

\begin{figure}
\includegraphics[width=85mm]{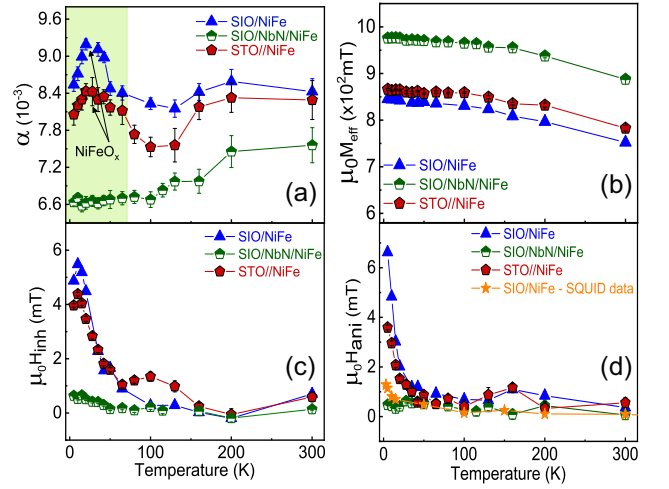}
\caption{\label{Fig3}BBFMR spectroscopic parameters as a function of temperature. (a) Gilbert damping $\alpha$, (b) effective magnetization ($M_\mathrm{eff}$), (c) inhomogeneous line broadening ($\mu_{0}H_\mathrm{inh}$), and (d) anisotropy field $\mu_0 H_\mathrm{ani}$ plotted as a function of temperature for SIO($5\;\mathrm{nm}$)/NiFe($5\;\mathrm{nm}$) (blue symbols), NiFe($5\;\mathrm{nm}$) (red symbols) SIO($5\;\mathrm{nm}$)/NbN($3\;\mathrm{nm}$)/NiFe($5\;\mathrm{nm}$) (green symbols) samples. For comparison, $\mu_0 H_\mathrm{c}$ derived from SQUID magnetization measurements is also shown in (d) for a SIO($5\;\mathrm{nm}$)/NiFe($3.5\;\mathrm{nm}$) (orange symbols).}
\end{figure}

To more systematically investigate the role of an interfacial NiFeO$_x$ layer, we investigated SIO ($5\;\mathrm{nm}$)/NiFe ($5\;\mathrm{nm}$) bilayer, NiFe ($5\;\mathrm{nm}$) single layer and SIO ($5\;\mathrm{nm}$)/NbN ($3\;\mathrm{nm}$)/NiFe($5\;\mathrm{nm}$) trilayer samples at $5\;\mathrm{K}\leq T \leq 300\;\mathrm{K}$ and extracted the FMR spectroscopic parameters. Fig.~\ref{Fig3}(a) shows the Gilbert damping $\alpha$ as a function of temperature for these three samples (Fitting procedures are described in supplementary material Fig. S2). For both the SIO/NiFe bilayer and the NiFe single layer samples, the Gilbert damping $\alpha$ increases at low temperatures, reaches a maximum around $25\;\mathrm{K}$, and then decreases with decreasing temperature, highlighted as a yellow shaded region in Fig.~\ref{Fig3}(a). In contrast, we only observe a weak temperature dependence for the SIO/NbN/NiFe trilayer sample, which is in accordance with earlier reports of elemental 3d-transition FMs.~\cite{25,26}  We attribute the observed maximum in Gilbert damping for the SIO/NiFe bilayer and NiFe single layer samples to the antiferromagnetic ordering\cite{14,41} of an oxide layer (thickness $\sim$0.5 nm) formed between SIO or STO and NiFe. This interfacial antiferromagnetic oxide layer also contributes to the damping due to magnetic fluctuations near the N\'{e}el temperature, which enhances the spin mixing conductance across the interface and thus increases the observed $\alpha$. From additional temperature-dependent BBFMR measurements on a SIO($5\;\mathrm{nm}$)/NiFe($7\;\mathrm{nm}$) bilayer, we extract an estimate for $g^{{\uparrow\downarrow}}_\mathrm{eff}(T)$ and find an enhancement of $g^{{\uparrow\downarrow}}_\mathrm{eff}$ around $50\;\mathrm{K}$ (see supplementary material). Similar results have been reported by L. Frangou \textit{et al.}\cite{14}, where they showed that the contribution of an interfacial antiferromagnetic oxide layer formed between SiO$_2$ and NiFe manifests as a peak in $\alpha$ near $T\sim 50\;\mathrm{K}$. At these low temperatures, we also find a larger Gilbert damping for the SIO/NiFe bilayer sample compared to the NiFe single layer sample. We attribute this observation to the fact that the SIO thin film has larger roughness than the STO substrate, which promotes a higher amount of NiFe oxidation. The effect of NiFe oxidation also manifests itself in the values extracted for $M_\mathrm{eff}$, as plotted in Fig.~\ref{Fig3}(b) (see also Fig. S3 in the supplementary material). For the SIO/NiFe bilayer and NiFe single layer samples, the extracted $M_\mathrm{eff}$ is significantly lower as for the SIO/NbN/NiFe trilayer sample. This change in $M_\mathrm{eff}$ may be attributed to an additional surface anisotropy, which originates from the formation of the NiFeO$_x$ layer or the direct contact of NiFe to the SIO (see supplementary material for further discussion). Most interestingly, we also find a dramatic change in the temperature dependence for the SIO/NiFe bilayer, and NiFe single layer samples in $\mu_{0}H_\mathrm{inh}$ (Fig.~\ref{Fig3}(c)) and $\mu_0 H_\mathrm{ani}$ (Fig.~\ref{Fig3}(d)) as compared to the SIO/NbN/NiFe trilayer sample. We can rule out the enhancement of two magnon scattering for the observed increase in $\mu_{0}H_\mathrm{inh}$, since we extract similar values in out-of-plane BBFMR measurements (see supplementary material).

\begin{figure}
\includegraphics[width=85mm]{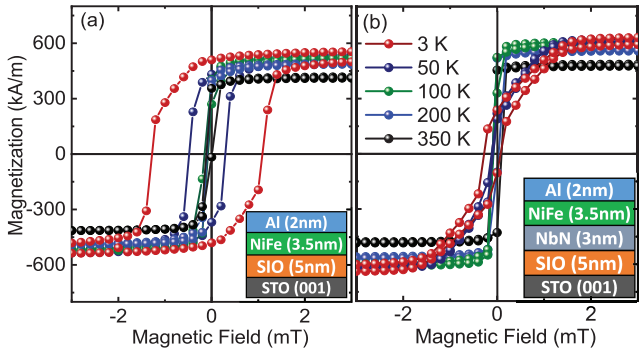}
\caption{\label{Fig4}SQUID magnetometry measurements: hysteresis loops for in-plane applied magnetic fields recorded at various temperatures for (a) SIO($5\;\mathrm{nm}$)/NiFe($3.5\;\mathrm{nm}$)/Al ($2\;\mathrm{nm}$ capping layer) and (b) SIO($5\;\mathrm{nm}$)/NbN($3\;\mathrm{nm}$)/NiFe($3.5\;\mathrm{nm}$)/Al ($2\;\mathrm{nm}$ capping layer), (stack illustrations are shown respectively as inset).}
\end{figure}

To further investigate the role of the interface oxide layer, we conducted magnetization measurements on SIO ($5\;\mathrm{nm}$)/NiFe ($3.5\;\mathrm{nm}$) bilayer and SIO ($5\;\mathrm{nm}$)/NbN ($3\;\mathrm{nm}$)/NiFe ($3.5\;\mathrm{nm}$) trilayer samples. We reduced the thickness of the NiFe as compared to the samples above used for the temperature dependent BBFMR studies to enhance the contributions from the thin oxidized NiFe interfacial layer. The magnetization measurements for the SIO/NiFe bilayer and the SIO/NbN/NiFe trilayer sample are shown in Fig.~\ref{Fig4} (a) and (b), respectively. As evident from these magnetization versus field measurements, we find a reduction by 15\% in the saturation magnetization (extracted by using the nominal thickness of the NiFe layer) for the SIO/NiFe bilayer as compared to the SIO/NbN/NiFe trilayer. If we assume this reduction originates only from oxidization, the NiFeO$_x$ thickness is $\sim0.5\;\mathrm{nm}$. However, even for the SIO/NbN/NiFe trilayer sample we find a much lower saturation magnetization than for bulk NiFe, which we attribute to a combination of a magnetic dead layer and experimental uncertainties in the determination of the volume of the NiFe layer. Moreover, we observe a strong increase in coercive field $\mu_0 H_\mathrm{c}$ for the SIO/NiFe bilayer sample as compared to the SIO/NbN/NiFe trilayer sample for temperatures below $100\;\mathrm{K}$. We attribute these larger $\mu_0 H_\mathrm{c}$ values to an exchange bias effect, where the ferromagnetic domains of NiFe are pinned by the antiferromagnetic NiFeO$_x$ phase.\cite{32,33}
We note that we do not observe a significant and systematic dependence of $\mu_0 H_\mathrm{c}$ on the external magnetic field applied while cooling down the sample. In addition, we extract a similar temperature dependence of the coercive field as compared to the determined BBFMR parameters for the SIO/NiFe bilayer sample. To illustrate this, we plotted the coercive field $\mu_0 H_\mathrm{c}$ in Fig.~\ref{Fig3}(d) as green-shaded symbols. A similar temperature dependence is found for the BBFMR extracted $\mu_0 H_\mathrm{ani}$ parameter and $\mu_0 H_\mathrm{c}$, indicating the same physical origin of both phenomena, \textit{i.e.,} an oxide layer formed at the SIO/NiFe interface.

In summary, the magnetization dynamics parameters of SIO/NiFe bilayers were studied as a function of temperature using the BBFMR technique. The room temperature measurements show spin pumping from NiFe into SIO, which is systematically studied and quantified by investigating a thickness series of the NiFe layer. The extracted spin mixing conductance of SIO/NiFe bilayers agrees well with results for NiFe/heavy metal materials and proves the potential application of SIO for SOT devices. In our low temperature BBFMR measurements, we find a significant enhancement of the Gilbert damping parameter around $50\;\mathrm{K}$. We attribute this observation to the formation of an oxide layer between NiFe and SIO, which orders antiferromagnetically at $50\;\mathrm{K}$ and thus leads to an enhancement of the spin mixing conductance via magnetic fluctuations.~\cite{27}
Moreover, $\mu_0 H_\mathrm{inh}$ and $\mu_0 H_\mathrm{ani}$ exhibit an increase at low temperatures. We compared these results to a SIO/NbN/NiFe trilayer sample, and found that the formation of the oxidation layer can be avoided by inserting a thin NbN spacer. Additional magnetization data showed an exchange bias effect between NiFe and the antiferromagnetic oxide layer and a reduction in the saturation magnetization for the bilayer. Our work provides a new perspective on spin current transport across metallic ferromagnet/SOT-active oxide interfaces. In particular, our results show that a NiFeO$_x$ layer at the interface of NiFe/SIO heterostructures leads to an enhanced spin pumping at room temperature. This enhancement can be either attributed to an enhanced spin mixing conductance or an increase in spin memory loss mediated by the NiFeO$_x$ layer. Further studies are required to analyze, whether this interfacial oxide layer is detrimental or beneficial for the spin current transport across the SIO/NiFe interface.

\section*{Supplementary Material}
See supplementary material for details on the growth parameters and supporting BBFMR data.

\begin{acknowledgements}
We acknowledge financial support by the German Academic Exchange Service (DAAD) via project no.~57452943 and by the DFG via project AL 2110/2-1. T.S.S and M.S.R acknowledge funding from DST 
(PH1920541DSTX002720), and grant SR/NM/NAT/02-2005. M.S.R. and K.S. acknowledge funding from SERB (EMR/2017/002328).
\end{acknowledgements}

\section*{AIP Publishing Data Sharing Policy}
The data that support the findings of this study are available from the corresponding author upon reasonable request.

\nocite{*}
\bibliography{Bibliography}
\end{document}